\def\answ{b }
\input harvmac
\input labeldefs.tmp
\writedefs
\overfullrule=0pt

\input epsf
\def\fig#1#2#3{
\xdef#1{\the\figno}
\writedef{#1\leftbracket \the\figno}
\nobreak
\par\begingroup\parindent=0pt\leftskip=1cm\rightskip=1cm\parindent=0pt
\baselineskip=11pt
\midinsert
\centerline{#3}
\vskip 12pt
{\bf Fig. \the\figno:} #2\par
\endinsert\endgroup\par
\goodbreak
\global\advance\figno by1
}
\newwrite\tfile\global\newcount\tabno \global\tabno=1
\def\tab#1#2#3{
\xdef#1{\the\tabno}
\writedef{#1\leftbracket \the\tabno}
\nobreak
\par\begingroup\parindent=0pt\leftskip=1cm\rightskip=1cm\parindent=0pt
\baselineskip=11pt
\midinsert
\centerline{#3}
\vskip 12pt
{\bf Tab. \the\tabno:} #2\par
\endinsert\endgroup\par
\goodbreak
\global\advance\tabno by1
}

\def\d{{\rm d}}
\def\e#1{{\rm e}^{#1}}
\def\las{\{ \lambda_i\}}
\def\mus{\{ \mu _k \}}
%
\def\pre#1{ (preprint {\tt #1})}
%
%
%
\lref\ICK{A.G.~Izergin, D.A.~Coker and V.E.~Korepin,
{\it J. Phys.} A 25 (1992), 4315.}
\lref\Ize{A.G.~Izergin, {\it Sov. Phys. Dokl.} 32 (1987), 878.}
\lref\Suth{ B. Sutherland, {\it PRL} 19 (1967), 103.}
\lref\Kor{V.E. Korepin, {\it Commun. Math. Phys} 86 (1982), 391.}
\lref\Bax{R.J.~Baxter, {\sl Exactly Solved Models in
Statistical Mechanics} (San Diego, CA: Academic).}
\lref\Kup{G.~Kuperberg, 
{\it Internat. Math. Res. Notices} 3 (1996), 139\pre{math.CO/9712207}.}
\lref\Sog{K.~Sogo, {\it Journ. Phys. Soc. Japan}
62, 6 (1993), 1887.}
\lref\Hir{R. Hirota, {\it Journ. Phys. Soc. Japan} 56 (1987), 4285.}
\lref\Wieg {O. Lipan, P.B. Wiegmann, A. Zabrodin,  preprint {\tt solv-int/9704015}.}
\lref\WZ {P.B. Wiegmann, A. Zabrodin, preprint {\tt hep-th/9909147}.}
\lref\WZK{I. Krichever, O. Lipan, P. Wiegmann and A. Zabrodin, preprint {\tt hep-th/9604080}.}
\lref\UT{K.~Ueno and K.~Takasaki, {\it Adv. Studies in Pure Math.} 4 (1984), 1.}
\lref\AvM{M.~Adler and P.~van~Moerbeke, {\it Duke Math. Journal}
80 (1995), 863\pre{solv-int/9706010};
preprint {\tt math.CO/9912143}.}
\lref\To{M. Toda, {\it Journ. Phys. Soc. Japan} 22 (1967), 431;
{\it Prog. Theor. Phys. Suppl.} 45 (1970), 174.}
\lref\Fla{H. Flaschka, {\it Phys. Rev.} B 9 (1974); {\it Prog. Theor.
Phys.} 51 (1974), 703.}
\lref\GMMMO{A.~Gerasimov, A.~Marshakov, A.~Mironov, A.~Morozov
and A.~Orlov, {\it Nucl. Phys.} B 357 (1991), 565.}
\lref\L{E.~Lieb, {\it Phys. Rev. Lett. } 18 (1967), 692.}
\lref\Li{E.~Lieb, {\it Phys. Rev. Lett.} 18 (1967), 1046.}
\lref\Lie{E.~Lieb, {\it Phys. Rev. Lett.} 19 (1967), 108.}
\lref\Lieb{E.~Lieb,  {\it  Phys. Rev.} 162 (1967), 162.}
\lref\Kas{P.W.~Kasteleyn, {\it Physica} 27 (1960), 1209.}
\lref\Fi{M.E.~Fisher, {\it Phys. Rev.} 124 (1961), 1664.}
\lref\LW{W.T.~Lu and F.Y.~Wu, {\it Phys. lett. A } 259 (1999), 108.}
\lref\JPS{W.~Jockush, J.~Propp and P.~Shor, preprint {\tt math.CO/9801068}.}
%
\lref\CEP{H.~Cohn, N.~Elkies and J.~Propp, {\it Duke Math. Journal}
85 (1996), 117.}
%
\lref\EKLP{N.~Elkies, G.~Kuperberg, M.~Larsen and J.~Propp,
{\it Journal of Algebraic Combinatorics} 1 (1922), 111; 219.}
\lref\Br{D.M.~Bressoud, {\sl Proofs and Confirmations:
The Story of the Alternating Sign Matrix Conjecture},
Cambridge University Press, Cambridge, 1999}
\lref\BP{D.~Bressoud and J.~Propp, {\it Notices of the AMS} June/July (1999), 637.}
\lref\Deift{P.~Deift, T.~Kriecherbauer, K.T-R.~McLaughlin, S.~Venakides and 
X.~Zhou, {\it Commun. on Pure and Applied Math.} 52
(1999), 1491.}
%
\lref\BBOY{M.T.~Batchelor, R.J.~Baxter, M.J.~O'Rourke and C.M.~Yung,
{\it J. Phys.} A28 (1995), 2759.}
\lref\Ken{R.~Kenyon, {\sl The planar dimer model with boundary:
a survey}, preprint\hfil\break
({\tt http://topo.math.u-psud.fr/$\sim$kenyon/papers.html}).}
\Title{
\vbox{\baselineskip12pt\hbox{YITP-00-13}\hbox{{\tt cond-mat/0004250}}}}
{{\vbox {
\vskip-10mm
\centerline{\bf Thermodynamic limit of the Six-Vertex Model}
\vskip2pt
\centerline{\bf with Domain Wall Boundary Conditions}
}}}
\medskip
\centerline{V.~Korepin {\it and} P.~Zinn-Justin}\medskip
\centerline{\sl C.N.~Yang Institute for Theoretical Physics}
\centerline{\sl State University of New York at Stony Brook}
\centerline{\sl Stony Brook, NY 11794--3840, USA}
\vskip .2in
\noindent 
We address the question of the dependence of the bulk free energy on boundary 
conditions for the six vertex model.
Here we compare the bulk free energy for periodic and domain wall 
boundary conditions.
Using a determinant representation for the partition function with domain
wall boundary conditions, we 
derive Toda differential equations and solve them asymptotically in order
to extract the bulk free energy. We find that it is different and bears no
simple relation with the free energy for periodic boundary conditions.
The six vertex model with domain wall boundary conditions is 
closely related to algebraic combinatorics (alternating sign matrices).
This implies new results for the weighted counting for large size
alternating sign matrices. Finally we comment on the interpretation of
our results, in particular in connection with domino tilings (dimers on a
square lattice).
\Date{04/2000}

\newsec{Introduction}

The six vertex model is an important  model of classical statistical
mechanics in two dimensions. The prototypical model is the ice model, which
was solved by Lieb \L\ in 1967
by means of Bethe Ansatz, followed  by several 
generalizations \Li, \Lie, \Lieb.
The solution of the most general six vertex model was given
by Sutherland \Suth\ in 1967.
The bulk free energy was calculated in these papers for periodic boundary 
conditions (PBC). 
A detailed classification of the phases of the model can be found 
for example in the book \Bax\ (see also the more recent
work \BBOY\ on anti-periodic boundary conditions).\par

Earlier, in 1961 Kasteleyn, while studying dimer arrangements
on a quadratic lattice, expressed doubts on the independence of the bulk free
energy on boundary conditions \Kas.
For more on dimer arrangements,
see \Kas, \Fi\ and \LW. Interest in this subject was renewed with recent work
on domino tilings (which are equivalent to dimers on a square lattice)
of an Aztec diamond \JPS, \CEP, demonstrating a strong effect of the boundary
on a typical domino configuration (see also \Ken).
Dimers (or domino tilings) can be considered as a particular case of
the six vertex model, and therefore a natural question is to investigate
the effect of boundary conditions on the thermodynamic limit of the six vertex
model.

Independently of this, new boundary conditions of the six-vertex model,
the so-called domain wall boundary conditions (DWBC),
were first introduced in 1982 \Kor\ (we shall define them
in detail below). An important recursion relation for the 
partition function was discovered in this paper. Later these recursion 
relations helped to find a determinant representation for the partition function
of the six vertex model with DWBC \Ize, \ICK.
The determinant representation simplifies somewhat in the homogeneous case.
In this case the partition function satisfy Toda
differential equation \Sog. In this paper we use this 
differential equation in order to calculate the bulk free energy for
domain wall boundary conditions. 

Let us all mention that there is a one to one correspondence between 
arrow configurations in the six vertex model with DWBC
and Alternating Sign Matrices (ASM) \EKLP. This mapping was 
used in order to count the number of ASM.
More on ASM can be found in \Br\ and \BP.

The plan of the paper is as follows.
In Section 2 we define the six-vertex model with domain wall boundary 
conditions, and
derive the determinant representation for the partition function. In Section
3 we derive Toda differential equation for the partition function. 
In Section 4 we consider the thermodynamic
limit; we derive the explicit expression of the bulk free energy in the 
ferroelectric and disordered phases,
and compare it with PBC.
Finally, in Section 5 we comment on the connection of our results
with other subjects (ASM, domino tilings, height model) 
and conclude this discussion in Section 6.

\newsec{Determinant representation of the partition function
of the six-vertex model}
In this section we shall define the {\it inhomogeneous} six-vertex model
with domain wall boundary conditions, and rewrite its partition
function as a determinant. We will then particularize our formula
to the homogeneous case.

\fig\conf{A configuration of the inhomogeneous
six-vertex model with domain wall boundary 
conditions.}{\epsfxsize=4.5cm\epsfbox{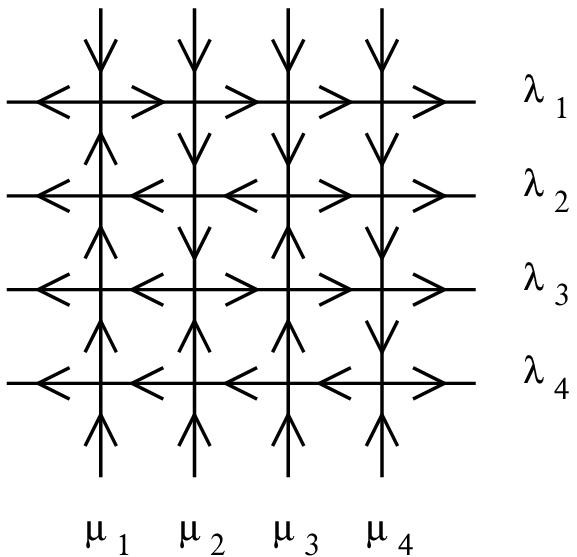}}
First we define the configurations of the model. They are given
by assigning arrows to each edge of a $N\times N$ square lattice
(see Fig.~\conf). The ``domain wall'' boundary conditions correspond
to fixing the horizontal external arrows to be outgoing and
the vertical external arrows to be incoming. The partition function
is then obtained by summing over all possible configurations:
\eqn\Z{
Z=\sum_{\hbox{arrow configurations}}\ \prod_{i,k=1}^N w_{ik}
}
where the statistical weights $w_{ik}$ are assigned to each {\it vertex}
of the lattice.
Since we are considering an inhomogeneous model, we need two
sets of spectral parameters $\las$ and $\mus$ which
are associated with the horizontal and vertical lines.
The weight $w_{ik}$ depends on the arrow configuration around
the vertex $(i,k)$ and is given by
\eqn\wei{
w_{ik}=\cases{
a(\lambda_i,\mu _k)\qquad \hbox{\raise-6mm\hbox{\epsfxsize=3.5cm\epsfbox{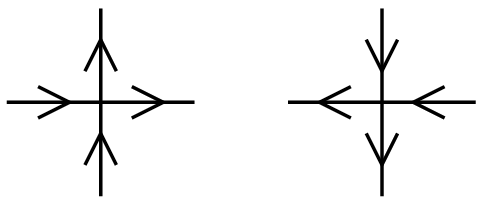}}}\cr
b(\lambda_i,\mu _k)\qquad \hbox{\raise-6mm\hbox{\epsfxsize=3.5cm\epsfbox{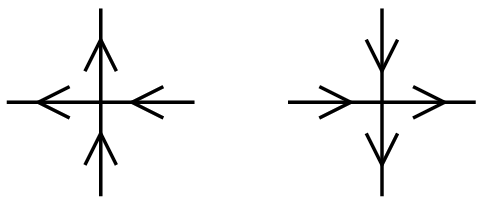}}}\cr
c(\lambda_i,\mu _k)\qquad \hbox{\raise-6mm\hbox{\epsfxsize=3.5cm\epsfbox{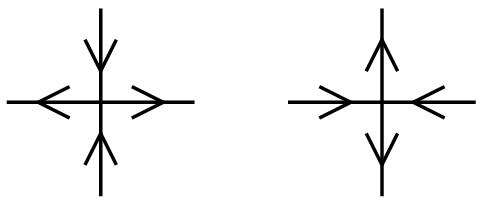}}}\cr
}
}
(all other weights are zero)
where the functions $a$, $b$, $c$ are chosen as follows:
\eqn\weib{
\eqalign{
a(\lambda,\mu )&=\sinh(\lambda-\mu -\gamma)\cr
b(\lambda,\mu )&=\sinh(\lambda-\mu +\gamma)\cr
c(\lambda,\mu )&=\sinh(2\gamma)\cr
}}
Here $\gamma$ is an anisotropy parameter which does not depend on the
lattice site. The partition function is therefore a function of
the $2N$ spectral parameters and we shall denote it by
$Z_N(\las,\mus)$.

The model thus defined satisfies the following essential property
(Yang--Baxter equation) shown on Fig.~\ybe. 
The vertex with diagonal edges is assigned
weights (the so-called $R$ matrix) which are the same
as the usual weights, up to a shift of the difference of the spectral
parameter. Here we shall not need
the explicit expression of the $R$ matrix.
\fig\ybe{Yang--Baxter equation. Summation over arrows of the {\it internal}
edges is implied, whereas external arrows are 
fixed.}{\epsfxsize=4.5cm\epsfbox{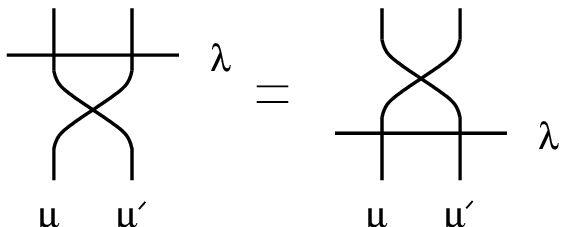}}

We shall now list the following four properties which
determine entirely $Z_N(\las,\mus)$ and sketch
their proof (for a detailed algebraic proof
the reader is referred to \ICK):

\item{a)} $Z_1=\sinh(2\gamma)$.

This is by definition.

\item{b)} $Z_N(\las,\mus)$ is a symmetric function of the
$\las$ and of the $\mus$.

It is sufficient to prove that exchange of $\mu _i$ and $\mu _{i+1}$
(for any $i$) leaves the partition function unchanged.
This can be obtained by repeated use of the Yang--Baxter property:
\eqn\permut{
\eqalign{
R_{\downarrow\downarrow}(\mu _i-\mu _{i+1}) &Z_N(\{\ldots\mu _i,
\mu _{i+1}\ldots\})
=\hbox{\raise-1.7cm\hbox{\epsfxsize=2.8cm\epsfbox{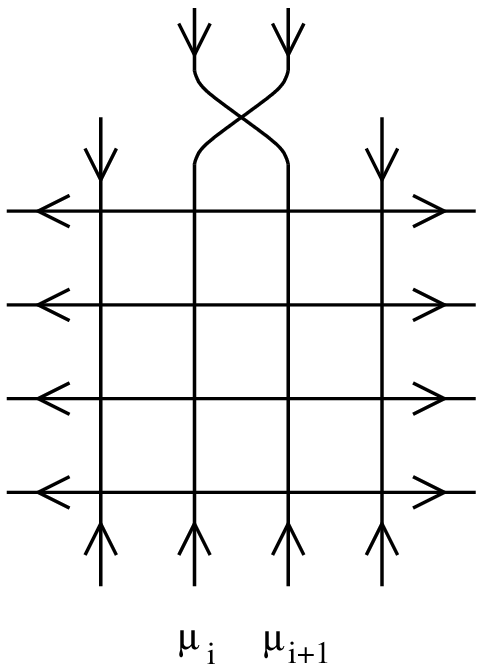}}}
=\hbox{\raise-1.7cm\hbox{\epsfxsize=2.8cm\epsfbox{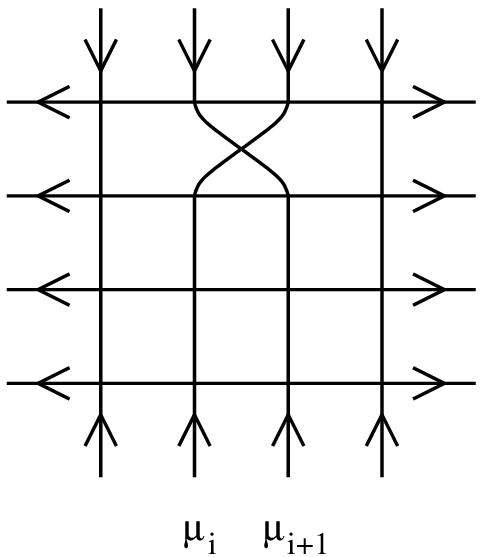}}}\cr
=\cdots
&=\hbox{\raise-2.4cm\hbox{\epsfxsize=2.8cm\epsfbox{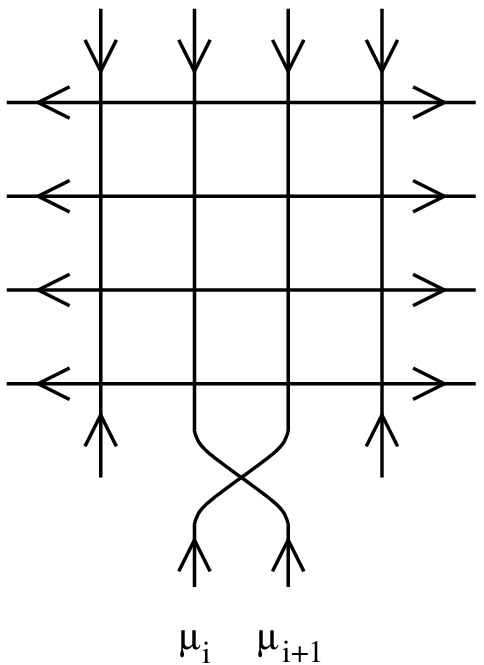}}}
=R_{\uparrow\uparrow}(\mu _i-\mu _{i+1}) Z_N(\{\ldots\mu _{i+1},
\mu _i\ldots\})
\cr
}}
where $R_{\uparrow\uparrow}=R_{\downarrow\downarrow}$ is the
appropriate entry of the $R$ matrix;
and similarly for the $\las$.

\item{c)} $Z_N(\las,\mus)=\e{-(N-1)\lambda_i}P_{N-1}(\e{2\lambda_i})$
where $P_{N-1}$ is a polynomial of degree $N-1$, and similarly for
the $\mu _k$.

Let us choose one configuration. Then the only weights
which depend on $\lambda_i$ are the $N$ weights on row $i$.
Since the outgoing arrows are in opposite directions, at least one
of the weights must be $c$. Therefore there are at most $N-1$ weights
$a$ and $b$, and the product of all weights is of the form
$\e{-(N-1)\lambda_i}P_{N-1}(\e{2\lambda_i})$. This property remains
of course valid when we sum over all configurations.

\item{d)} $Z_N(\las,\mus)$ obeys the following recursion relation:
\eqn\recur{
\eqalign{
Z_N(\las,\mus)_{\left|\lambda_j-\mu _l=\gamma\right.}
=&\sinh(2\gamma)
\prod_{\scriptstyle 1\le k\le N\atop\scriptstyle k\ne l} \sinh(\lambda_j-\mu _k+\gamma)\cr
&\prod_{\scriptstyle 1\le i\le N\atop\scriptstyle i\ne j} \sinh(\lambda_i-\mu _l+\gamma)\ 
Z_{N-1}(\las_{i\ne j},\mus_{k\ne l})}}

Because of property b), we can assume that $j=l=1$.
Since $\lambda_k-\mu _l=\gamma$ implies $a(\lambda_j-\mu _l)=0$,
by inspection all configurations with non-zero weights are of the
form shown on Fig.~\recurs. This immediately proves Eq.~\recur.
\fig\recurs{Graphical proof of the recursion relation.}{
\epsfxsize=4.5cm\epsfbox{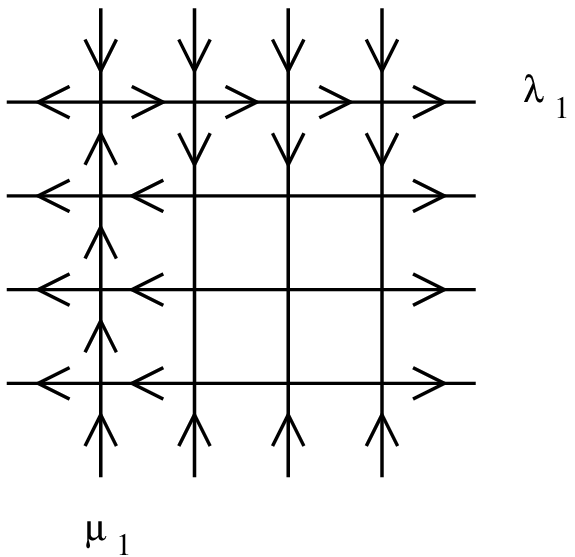}}

It is easy to see that the four 
properties a), b), c) and d) characterize entirely
$Z_N(\las,\mus)$. This is enough to prove that $Z_N(\las,\mus)$ has
the following determinant representation \ICK:
\eqn\detrep{\eqalign{
Z_N(\las,\mus)=&{\prod_{1\le i,k\le N}\sinh(\lambda_i-\mu _k+\gamma)
\sinh(\lambda_i-\mu _k-\gamma)
\over\prod_{1\le i<j\le N} \sinh(\lambda_i-\lambda_j)
\prod_{1\le k<l\le N} \sinh(\mu _k-\mu _l)}\cr
&\det_{1\le i,k\le N}
\left[{\sinh(2\gamma)\over\sinh(\lambda_i-\mu _k+\gamma)
\sinh(\lambda_i-\mu _k-\gamma)}\right]\cr}
}
Indeed, one can check that this expression satisfies the 
four properties listed above.

The expression \detrep\ might seem singular when two
spectral parameters $\lambda_i$ and $\lambda_j$ coincide (and
similarly for the $\mu _k$); but in fact the pole created
by the factor $\sinh(\lambda_i-\lambda_j)$ is compensated by
the zero of the determinant due to the fact
that two rows are identical. Therefore, particular care
must be taken when considering the homogeneous limit where
all the $\lambda_i$ are equal (and all the $\mu _k$). This limit
was studied in detail in \ICK, and we shall simply
summarize the result of the calculation. Let us call $t$ the common
value of $\lambda_i-\mu _k$ for all $i$ and $k$.
When the $\lambda_i$ are close to one
another one must Taylor expand the function
\eqn\defphi{
\phi(t)\equiv{\sinh(2\gamma)\over\sinh(t+\gamma)\sinh(t-\gamma)}
}
which appears in the determinant. This leads to the following expression:
\eqn\detrepb{
Z_N(t)={(\sinh(t+\gamma)\sinh(t-\gamma))^{N^2}
\over\left(\prod_{n=0}^{N-1} n!\right)^2} \det_{1\le i,k\le N}
\left[{\d^{i+k-2}\over\d t^{i+k-2}}\phi(t)\right]
}

\newsec{Determinant representation and Toda chain hierarchy}
We shall now investigate the properties of the determinant which
appears in Eq.~\detrepb, and for which we introduce the notation
\eqn\deftau{
\tau_N(t)=\det_{1\le i,k\le N} 
\left[m_{i+k-2}\right]}
with
\eqn\defm{
m_n={\d^n\over\d t^n}\phi(t)
}

Let us write
down the bilinear Hirota equation satisfied by the $\tau_N$. For
completeness, we recall that they are a consequence of
Jacobi's determinant identity:
\eqn\detid{
\hbox{\raise-0.7cm\hbox{\epsfxsize=1.5cm\epsfbox{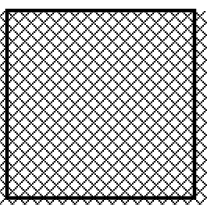}}}
\ 
\hbox{\raise-0.7cm\hbox{\epsfxsize=1.5cm\epsfbox{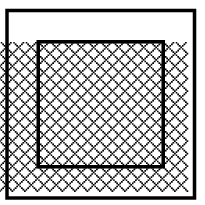}}}
=
\hbox{\raise-0.7cm\hbox{\epsfxsize=1.5cm\epsfbox{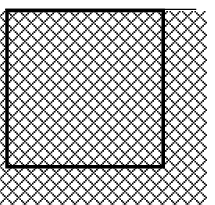}}}
\ 
\hbox{\raise-0.7cm\hbox{\epsfxsize=1.5cm\epsfbox{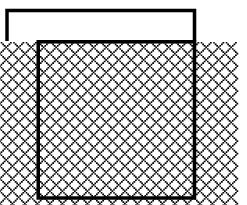}}}
-
\hbox{\raise-0.7cm\hbox{\epsfxsize=1.5cm\epsfbox{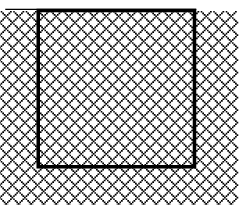}}}
\ 
\hbox{\raise-0.7cm\hbox{\epsfxsize=1.5cm\epsfbox{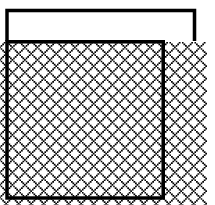}}}
}
The large squares represent a given matrix, and
the shaded regions are the sub-matrices
whose determinants one must consider.
Applying it to $\tau_{N+1}$ (up to a re-shuffling of the rows and columns),
we find \Sog:
\eqn\hirota{
\tau_N\tau_N''-\tau_N'{}^2=\tau_{N+1}\tau_{N-1}\qquad \forall N\ge 1
}
where primes denote differentiation with respect to $t$.
This is supplemented by the initial data: $\tau_0=1$ and $\tau_1=\phi$.
Equivalently, we have:
\eqn\hirotab{
(\log\tau_N)''={\tau_{N+1}\tau_{N-1}\over\tau_N^2}\qquad \forall N\ge 1
}
which is the form of the equation that we shall use.

Note that if we introduce the combinations $\e{\varphi_N}=\tau_N/\tau_{N-1}$,
$N\ge 1$,
Eq.~\hirotab\ implies for the $\varphi_N$:
\eqn\toda{
\varphi_N''=\e{\varphi_{N+1}-\varphi_N}-\e{\varphi_N-\varphi_{N-1}}
\qquad\forall N\ge 2
}
and $\varphi_1''=\e{\varphi_2-\varphi_1}$.
These are the usual Toda (semi-infinite) chain equations \To, \Hir, \Fla. 
Another possible
form is
\eqn\todab{
\psi_N''=-\sum_M C_{MN}\,\e{\psi_M} \qquad N\ge 1
}
with $\psi_N=\varphi_{N+1}-\varphi_N$ and $C_{MN}$ ($M$, $N\ge 1$) the
Cartan matrix of the semi-infinite diagram $A_\infty$.

This suggests a connection with the Toda chain hierarchy \UT, \AvM, \Wieg,
\WZ, \WZK, \GMMMO.
Indeed, let us mention that given a H{\"a}nkel matrix $(m_{i+k-2})$ -- 
i.e.\ whose entries only depend on $i+k$ -- 
the $m_n$ can be made to depend
on a set of parameters $\{ t_q\}_{q\ge 1}$ in such a way that the 
determinants $\tau_N$ 
become $\tau$-functions of the whole Toda (semi-infinite)
chain hierarchy \AvM. Namely, one must choose
\eqn\deftq{
m_n(\{ t_q\})=\int \d\rho(x)\, x^n\,\e{\sum_{q\ge 1} t_q x^q}
}
where $\d\rho(x)$ is an arbitrary measure\foot{This must be
considered as a formal expression; e.g.\ the measure may not
necessarily be smooth or positive.} (in the matrix model context \AvM, the
$t_q$ are the coefficients of the polynomial potential).
Here, we are in the simplest
situation where only one parameter $t_1\equiv t$ is allowed to evolve.
We immediately check that Eq.~\deftq\ implies that
$m_n(t)={\d^n\over\d t^n}m_0(t)$, which is consistent with Eq.~\defm.

\newsec{The thermodynamic limit}
We shall now consider the thermodynamic (i.e.\ large $N$) limit
of the expression \detrepb\ in the various regimes of the six-vertex model.
For that we shall use the Hirota equation in its form \hirotab.

When $N\to\infty$ it is expected that the partition function
behaves in the following way:
\eqn\asyZ{
\log Z_N(t)=-N^2 F(t) + O(N)
}
where $F(t)$ is the {\it bulk free energy} (we shall always
set the temperature $k_B T=1$). Our main goal is to
compute explicitly $F(t)$.

Comparing the expected asymptotic \asyZ\ with the exact formula
\detrepb, we find that the determinant $\tau_N$ must be of the form
\eqn\asytau{
\tau_N=\left(\prod_{n=0}^{N-1} n!\right)^2
\e{N^2 f(t)+O(N)}
}
where
\eqn\deff{
f(t)=-F(t)-\log(\sinh(t+\gamma)\sinh(t-\gamma))
}
We now want to substitute the expansion \asytau\ into the equation \hirotab.
For that we need to assume that the sub-dominant corrections to
the bulk free energy vary slowly as a function of $N$; we shall
discuss the validity of this assumption below. 
We then find that the expansion is consistent since both left and right
hand sides of \hirotab\ turn out to be of order $N^2$. The resulting equation
for $f$ is:
\eqn\diff{
f''=\e{2f}
}
This is an ordinary second order differential equation, which can be
readily solved. The general solution depends on two parameters
$\alpha$ and $t_0$:
\eqn\soldif{
\e{f(t)}={\alpha\over\sinh(\alpha(t-t_0))}
}
If the weights are chosen to be real, then the free energy should be
real and this implies that $\alpha$ must be real or purely imaginary.

So far everything we have done was independent of the particular form
of the function $\phi(t)$ and therefore independent of $\gamma$.
In order to fix the two constants in \soldif, we must now discuss
separately the different regimes of the six-vertex model.
Let us recall that the latter are usually distinguished by the
value of the parameter (cf Eq.~(8.3.21) of \Bax)
\eqn\defDel{
\Delta={a^2+b^2-c^2\over 2 ab}
}
The weights $a$, $b$, $c$ were defined in Eq.~\weib\
(with $\lambda-\mu \equiv t$). In this parameterization,
\eqn\defDelb{
\Delta=\cosh(2\gamma)}

\subsec{Ferro-electric phase: $\Delta>1$}
This corresponds to the parameters $\gamma$ and $t$ real;
we recall that the weights are given by
\eqn\weic{
a=\sinh(t-\gamma)\qquad
b=\sinh(t+\gamma)\qquad
c=\sinh(2\gamma)
}
with $|\gamma|<t$.
This is the so-called ferroelectric phase. In the
case of periodic boundary conditions, it is known that the
system is frozen in its ground state configuration, in which
all arrows are aligned: if $a>b$ all arrows point
up and to the right or down and to the left, whereas
if $b>a$ they point up and to the left or down and to the right.
The domain wall boundary conditions do not allow
all arrows to be aligned: the ground state will instead take the form
of Fig.~\gs. However at leading order in the large $N$ limit,
this does not affect the free energy, and we expect to find
the same result as for periodic boundary conditions.
\fig\gs{Ground state configuration of the ferroelectric phase
(for $b>a$; the case $a>b$ is obtained by taking the mirror image).}{\epsfxsize=3.5cm\epsfbox{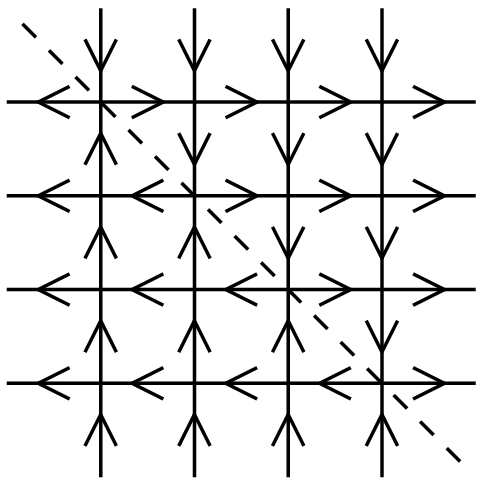}}

Indeed, it is easy to see that the relevant solution of Eq.~\diff\ is
\eqn\soldifb{
\e{f(t)}={1\over\sinh(t-|\gamma|)}}
and therefore the bulk free energy takes the form
\eqn\freea{
\e{-F(t)}=\sinh(t+|\gamma|)=\max(a,b)
}
in agreement with the case of periodic boundary conditions.

\subsec{Disordered phase: $-1<\Delta<1$}
In this regime, it is customary to make
the following redefinitions: 
\eqn\redefc{
\gamma'=i(\gamma+i\pi/2)
}
\eqn\redefb{
t'=i(t+i\pi/2)
}
and divide all the weights by $i$, so that they take the form:
\eqn\weie{
a=\sin(\gamma-t)\qquad
b=\sin(\gamma+t)\qquad
c=\sin(2\gamma)
}
and $\Delta=-\cos(2\gamma)$.
Using symmetry considerations, one can always assume $0<\gamma<\pi/2$.
We only consider the region $|t|<\gamma$ (where the weights are positive).

Taking into account these redefinitions,
the partition function becomes:
\eqn\detrepc{
Z_N(t)={(\sin(t+\gamma)\sin(t-\gamma))^{N^2}
\over\left(\prod_{n=0}^{N-1} n!\right)^2} \det_{1\le i,k\le N}
\left[{\d^{i+k-2}\over\d t^{i+k-2}}\phi(t)\right]
}
with a redefined $\phi(t)=\sin(2\gamma)/(\sin(t-\gamma)\sin(t+\gamma))$;
the determinant $\tau_N$ still satisfies Eq.~\hirota\ and
$f(t)$ defined by \deff\ is still a solution of Eq.~\diff.\foot{Note
that the sign is unchanged in Eq.~\diff; this is the combined
effect of the ``Wick rotation'' of $t$ ($t\to it$) and of
dividing all the weights by $i$ ($\e{f}\to -i\,\e{f}$).}

Let us mention that the partition function has been computed
exactly \Kup\ at three particular values of the parameters:
$t=0$, $\gamma=\pi/6$,
$\pi/4$ and $\pi/3$. In all three cases the expansion \asyZ\ and the
assumption of smoothness of the sub-dominant corrections (which is
necessary to derive the ordinary differential equation \diff) can
be checked exactly. We have also checked it numerically for a variety
of values of $t$ and $\gamma$.

We must now select the appropriate solution (of the form \soldif) of
the Eq.~\diff. Let us first assume that $|t|<\gamma$ (this is the only
physical region, i.e.\ where all the weights are positive). It is easy
to check that $f(t)$ must be an even function of $t$. The only even
solution of Eq.~\diff\ is
\eqn\soleven{
\e{f(t)}={\alpha\over\cos(\alpha t)}
}
where $\alpha$ remains to be determined. Note that this implies for $F(t)$
\eqn\freeb{
\e{-F(t)}=\sinh(\gamma-t)\sinh(\gamma+t){\alpha\over\cos(\alpha t)}
}

We must then use the boundary condition
given by $|t|=\pm \gamma$. At these values one can compute directly $Z_N(t)$.
Indeed the only non-zero configurations are of the form of Fig.~\gs, and 
we find
\eqn\freec{
Z_N(t=\pm \gamma)=\sin(2\gamma)^{N^2}
}
and therefore $\e{-F(t)}=\sin(2\gamma)$. Since the prefactor in \freeb\ 
vanishes when $|t|=\gamma$,
we conclude that $\alpha$ must be chosen in such a way that $\cos(\alpha t)$
is non-zero for $|t|<\gamma$, but vanishes as $|t|=\gamma$. This uniquely
determines $\alpha$ to be: $\alpha={\pi\over 2\gamma}$. We obtain the
final expression
\eqn\freed{
\e{-F(t)}=\sin(\gamma-t)\sin(\gamma+t){\pi/2\gamma\over \cos(\pi t/2\gamma)}
}
As a consistency check, one takes the limit $t\to\pm \gamma$ and finds
$\e{-F(t)}=\sin(2\gamma)$, as it should be. Also, note that for
$\gamma=\pi/4$, where the partition function is
known and independent of $t$, one finds indeed that $\e{-F(t)}=1$.

For further checks, let us set $t=0$; a more standard normalization of
the weights is then
\eqn\weid{
a=b=1\qquad c=2\cos\gamma
}
and the bulk free energy becomes
\eqn\freee{
\e{-F}={\pi\over 2} {\sin\gamma\over\gamma}
}
At $\gamma=\pi/6$, $\pi/4$, $\pi/3$, the values predicted by \freee\ coincide
with the large $N$ limit of the expressions of \Kup.
Also, this fits perfectly with some numerical computations of the
determinant we have performed.

We can compute the bulk energy (energy per unit site), which turns out to be
\eqn\ener{
E=(\cot\gamma-1/\gamma)\,\cot\gamma\,\log(2\cos\gamma)
}
Fig.~\monte\ shows the comparison with Monte-Carlo simulations.
The agreement is also very good.
\fig\monte{Energy $E$ as a function of the anisotropy $\Delta$. The curve
is given by Eq.~\ener, whereas the diamonds are the results
of Monte-Carlo simulations on 
lattices of size $N=64$.}{\epsfxsize=9cm\epsfbox{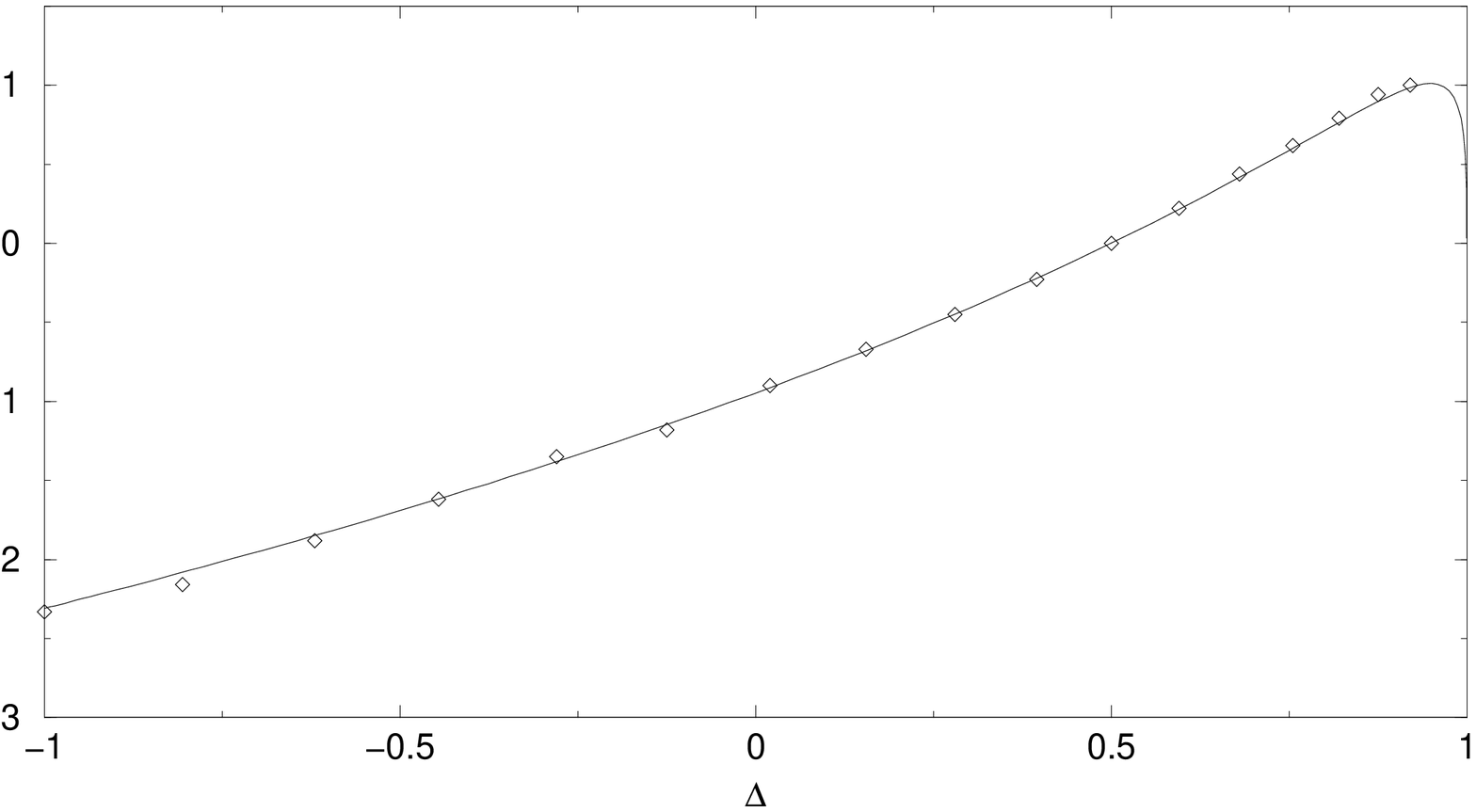}}

Finally let us mention that there seems to be no simple relation
between the PBC and DWBC bulk free energies: from an analytic point of view,
the DWBC free energy 
is an elementary function, whereas the PBC free energy is given by
a non-trivial integral. Furthermore, the DWBC free energy is always greater
then the PBC free energy, even at infinite temperature ($\Delta=1/2$),
see Fig.~\freepic.
\fig\freepic{Bulk free energies for PBC and DWBC as a function of
$\Delta$.}{\epsfxsize=8cm\epsfbox{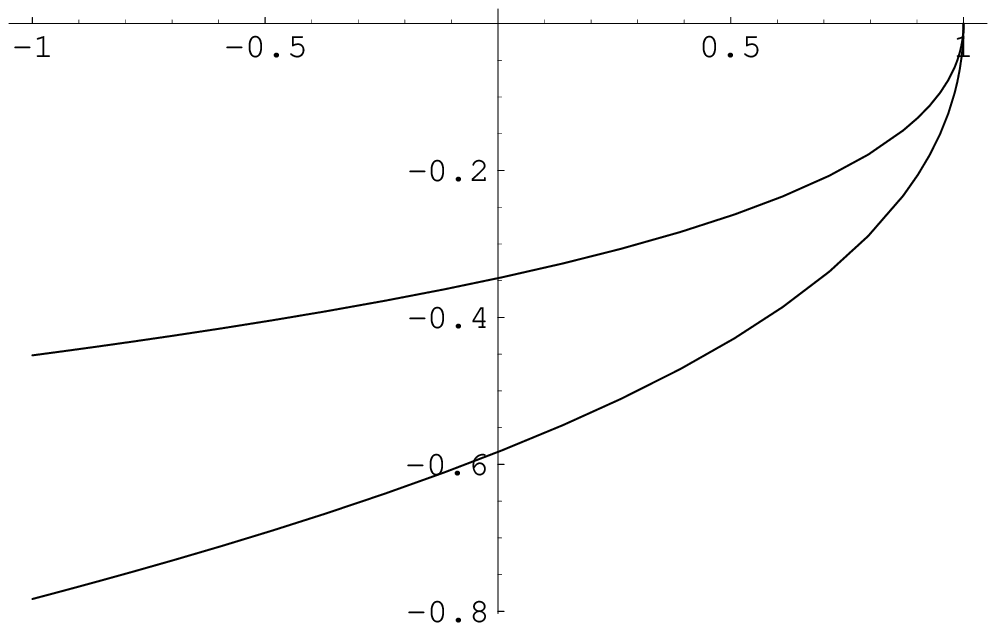}}

\subsec{Anti-ferroelectric phase: $\Delta<-1$}
In this phase, the smoothness assumption of the sub-dominant corrections
to the bulk free energy is {\it not} satisfied, as can be clearly
seen numerically. The ratio $Z_{N+1}Z_{N-1}/Z_N^2$ does not converge
in the large $N$ limit but instead has a pseudo-periodic behavior reminiscent
of the one-matrix model with several cuts \Deift, and slightly
more sophisticated methods are needed to analyze the large $N$ limit.
We leave this to a future publication.

\subsec{Phase transition at $\Delta=1$}
If the Boltzmann weights depend on a parameter (e.g.\ temperature),
it is known that with periodic boundary conditions, the system
undergoes phase transitions as $\Delta$ crosses $\pm 1$. 
Let us use the expressions of the bulk free energy found above
to clarify what happens in the case of domain wall boundary conditions.

Here we shall consider the transition from ferroelectric (low
temperature)
to disordered (high temperature)
regime, that is from $\Delta>1$ to $\Delta<1$.
The parameter that plays the role of deviation from criticality
$T-T_c$ can be defined as
\eqn\devia{
T-T_c\equiv 1-\Delta
}
We assume that $b>a$ (the case $a>b$ can be treated similarly),
and re-scale the weights so that $b=1$. With this convention,
we simply have in the ferroelectric phase:
\eqn\freeh{
\e{-F}=1 \qquad \Delta>1
}
(cf Eq.~\freea).
Let us now consider $\Delta\to 1^-$. The weights are
\eqn\weif{
a={\sin(\gamma-t)\over\sin(\gamma+t)}\qquad
b=1\qquad
c={\sin(2\gamma)\over\sin(\gamma+t)}
}
with $\gamma=\pi/2+\epsilon$, $t=\pi/2+\epsilon x$; $x$ must be kept
fixed as $\epsilon\to 0$. Note that $\Delta=\cos(2\epsilon)$, so that
\eqn\devib{
T-T_c\propto\epsilon^2
}

Expanding the expression \freed\ for the free energy, we obtain:
\eqn\freei{
\e{-F}=1-{2(x-1)^2\over 3\pi}\epsilon^3+O(\epsilon^4)
}
Comparing \freeh\ and \freei, we find
a {\it second order} phase transition, with a singular part
$(T-T_c)^{3/2}$ corresponding to a critical exponent
$\alpha=-1/2$.
This is to be contrasted with the first order phase transition
that occurs in the case of PBC. Let us
however emphasize that the difference of orders of the
phase transitions is not that significant,
since the phase transition
is somewhat special (in the case of PBC, the correlation length
jumps from zero for $\Delta>1$ to infinity for $\Delta<1$).

\newsec{Some equivalences}
We shall now review some alternative interpretations of the partition function
of the six-vertex model with domain wall boundary conditions; these
equivalent formulations will shed some light on the property of
dependance on boundary conditions that was found.

\subsec{Alternating sign matrices}
Six-vertex model arrow configurations
with domain wall boundary conditions on a $N\times N$ lattice
are in one-to-one correspondence with {\it alternating sign matrices} (ASM)
of size $N$,
that is square matrices with entries $0$ or $\pm 1$ such that each row
and column has an alternating sequence of $+1$ and $-1$ (zeroes excluded)
starting and ending with a $+1$. 
Recalling that there are 6 weights which we shall call $a_1$, $a_2$,
$b_1$, $b_2$, $c_1$, $c_2$ in the order shown in Eq.~\wei,
the correspondence goes as follows:
given a six-vertex configuration, assign a $0$ to
each vertex $a$ or $b$ and $+1$ (resp.\ $-1$) to each
vertex $c_1$ (resp.\ $c_2$). One can show that
this map is bijective, and therefore,
the number of ASM is exactly equal to the partition function considered
before with $a=b=c=1$. For our purposes, let us define a refined counting
of ASM ($x$-enumeration in the language of \Kup) by assigning a weight $x$ to
each entry $-1$ of the ASM. The resulting quantity $A(N,x)$ is still related to
the six-vertex model; indeed, one can easily show that
\eqn\asm{
A(N,x)=x^{-N/2} Z_N(a=b=1,c=\sqrt{x})
}
If $0\le x\le 4$, one can set $x=4\cos^2\gamma$
and the weights are of the form \weid. One can then prove (extending
slightly the asymptotic expansion found in {\it 4.2}) that
\eqn\asyasm{
\log A(N,x)=N^2\log\left[{\pi\over2}{\sin\gamma\over\gamma}\right]
-{N\over 2}\log x + O(\log N)\qquad \forall x\in [0,4]
}
Though this equivalence does not directly provide any useful insight
into the issue adressed in this paper, the result \asyasm\ itself
might be of some mathematical interest.

\subsec{Tilings of the Aztec diamond}
A more illuminating equivalence is
that of domino tilings (i.e.\ dimers on a square lattice
in a dual description) and the six-vertex model at $\Delta=0$
-- both models are well-known to describe essentially one Dirac fermion.
This is illustrated on Fig.~\domino. Since each vertex of type $c_1$
has $2$ possible corresponding domino tiling configurations,
one must assign it a Boltzmann weight of $2$ in order to count each
domino tiling exactly once; however with most boundary conditions
there are as many vertices of type $c_1$ and $c_2$, and therefore one can
give them both a weight of $\sqrt{2}$ instead, which leads to the values
$a=b=1$, $c=\sqrt{2}$ of the parameters.
\fig\domino{Correspondence between vertices
of the six-vertex model and small patches of a domino 
tiling.}{\epsfxsize=7cm\epsfbox{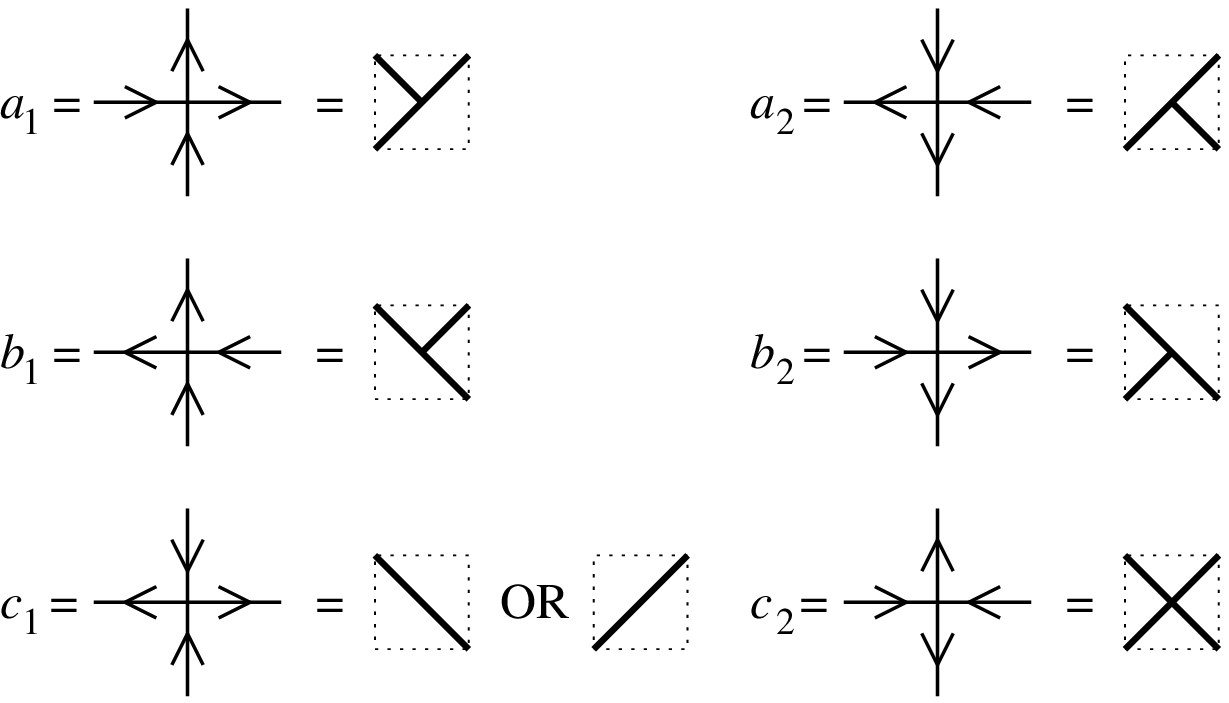}}
The more precise statement is that
the number of domino tilings of the {\it Aztec diamond} (see \JPS) is
equal (up to a small known prefactor) to the partition function
of the six-vertex model with domain wall boundary
conditions at $a=b=1$, $c=\sqrt{2}$, see Fig.~\dominob.
\fig\dominob{a) A configuration of the six-vertex model with DWBC,
and b) one possible corresponding tiling of the Aztec diamond.}
{\epsfxsize=7.5cm\epsfbox{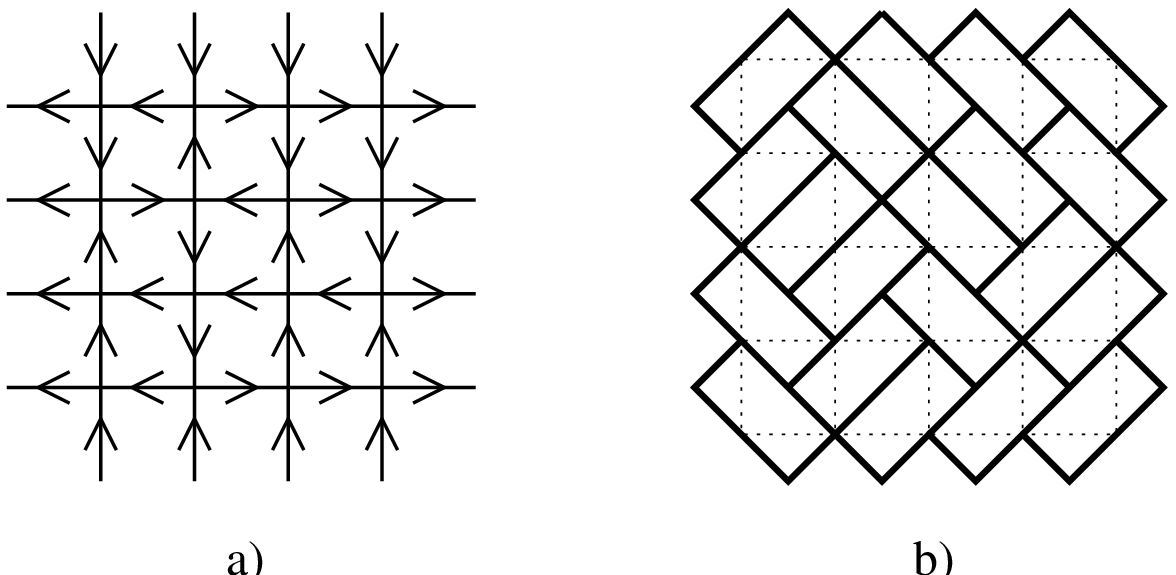}}
Since this a local correspondence of configurations it entends to all
correlation functions. Also, introducing some weights for the
local tiling patterns
amounts to changing the weights $a$, $b$, $c$, but always in such a way
that $\Delta$ remains zero.

These tilings have been
an object of interest for mathematicians, see in particular
\CEP, \EKLP. The ``arctic circle theorem'' \JPS\ shows that
as the size of the system grows large,
the domino configurations become frozen outside the circle
inscribed inside the diamond, and remain disordered but still
heterogeneous \CEP\ (i.e.\ non translationally invariant) inside the circle.
These statements have a straightforward equivalent in the six-vertex
language: we expect the one-point functions of the six-vertex model
at $\Delta=0$ with DWBC to be non-constant (following a similar
pattern as the tilings), and presumably a similar behavior at $\Delta\ne 0$.
This gives a qualitative understanding of the dependence
of the bulk free energy on the boundary conditions.

\subsec{Height model}
The general eight-vertex is well-known to be equivalent
to a class of height models (SOS/RSOS model). In
the case of the six-vertex model, there is a particularly simple
equivalence which goes as follows: given a six-vertex
configuration, integers are
assigned to each face of the lattice in such a way that going from
one face to a neighboring face, the number is increased by one if
the arrow in between goes right (and so, is decreased by one if it
goes left).
Conservation of the arrows ensures consistency of this procedure.
The Boltzman weights of the model are simply the weights
of the original six-vertex model expressed in terms of the new
height variables.

This equivalence is particularly interesting because it gives us a
simple intuitive explanation of the lack of thermodynamic limit due
to boundary conditions; this was studied in detail and proven
rigorously in the case of tilings ($\Delta=0$), see \Ken.
Let us consider the domain wall boundary conditions and translate these
into the language of our height model. They are fixed boundary conditions
for the heights, of the form:
$$\matrix{
0&1&\cdots&N-1&N\cr
1& &      &   &N-1\cr
\vdots&&  &   &\vdots\cr
N-1&&     &   &1\cr
N&N-1&\cdots&1&0\cr
}$$
where we have fixed arbitrarily the upper left height to be zero.

In the thermodynamic limit $N\to\infty$, let us define the rescaled
coordinates on the square lattice to be $x=k\,a$, $y=i\,a$ where
$a\equiv 1/N$ is the lattice spacing. The heights $h_{i,j}$ are
supposed to renormalize, according to standard lore, to a free massless
bosonic field:
\eqn\bos{
h_{i,j}\to \phi(x,y)
}
However, it is reasonable to assume
that in order to have a proper thermodynamic limit, the boundary conditions
must be well-defined in terms of the limiting field $\phi$. In the case
of DWBC, one finds that the boundary conditions become $\phi(x,0)=x/a$ etc,
which do not have a limit as $a\to 0$; in particular,
the variations of $\phi$ on the boundary diverge.
More generally, we can conjecture that only the boundary conditions
such that the variation of the function $\phi$ on the boundary can remain
bounded will lead to the usual thermodynamic limit. This is essentially
what is proven in \Ken\ in the case $\Delta=0$.

\newsec{Conclusion}
In this work, we have computed explicitly the large $N$ asymptotic
behavior
of a $N\times N$ determinant which plays the role of partition function
of the six-vertex model with domain wall boundary conditions. This gives
rise to particularly simple expressions for the bulk free energy of this
model (Eqs.~\freea\ and \freed). One important question is to
physically interpret the discrepancy of the bulk free energy found 
when comparing domain wall and periodic boundary conditions of
the six-vertex model, which is somewhat
contrary to standard lore on the thermodynamic
limit of statistical models. Some clues were given in the previous
section, where various equivalences were discussed. In particular it
was shown how ``generic'' fixed boundary conditions for the six-vertex
model do not lead to a well-defined thermodynamic limit. It would be
useful to make these arguments more rigorous. Also, it would be most
interesting to find a more quantitative description of the
non-translational invariance created by the boundary conditions,
and in particular to prove a
``generalized arctic circle theorem'' for any value of the
parameter $\Delta$ of the six-vertex model.

\bigskip
\centerline{\bf Acknowledgements}
It is a pleasure to acknowledge stimulating discussions with E. Lieb,
F.Y.~Wu,  S.~Ruijsenaars (who informed us that he had similar results) and 
R.~Behrends (who participated in the early stages of this project).
This work was supported by the National Science Foundation under
grant number PHY-9605226 (V.K.).
\footatend\vfill\supereject\immediate\closeout\rfile\writestoppt
\baselineskip=14pt\centerline{{\bf References}}\bigskip{\frenchspacing%
\parindent=20pt\escapechar=` \input refs.tmp\vfill\eject}\nonfrenchspacing
\bye